\documentclass[12pt]{article}
\usepackage{amsfonts}
\usepackage{amsmath,mathrsfs,bm,amssymb,color}
\usepackage[width=15cm,height=8.5in]{geometry}
\usepackage{hyperref}

\newtheorem{theorem}{Theorem}[section]

\newtheorem{lemma}[theorem]{Lemma}
\newtheorem{proposition}[theorem]{Proposition}
\newtheorem{corollary}[theorem]{Corollary}
\newtheorem{remark}[theorem]{Remark}

\newtheorem{assumption}[theorem]{Assumption}

\providecommand{\bc}[1]{\begin{corollary}\label{#1}}
\providecommand{\ec}{\end{corollary}}
\providecommand{\bp}[1]{\begin{proposition}\label{#1}}
\providecommand{\ep}{\end{proposition}}

\newenvironment{proof}[1][Proof]{{\sc #1.} }{\hfill $\Box$}

\newcommand{\bt}[1]{\begin{theorem}\label{#1}}
\newcommand{\et}{\end{theorem}}
\newcommand{\bl}[1]{\begin{lemma}\label{#1}}
\newcommand{\el}{\end{lemma}}

\newcommand{\br}[1]{\begin{remark}\label{#1}}
\newcommand{\er}{\end{remark}}

\makeatletter
\@addtoreset{equation}{section}
\makeatother

\def\bbbone{{\mathchoice {\rm 1\mskip-4mu l} {\rm 1\mskip-4mu l}
{\rm 1\mskip-4.5mu l} {\rm 1\mskip-5mu l}}}
\def\one{\bbbone}

\newcommand{\kp}{\one_\Lambda^\perp(k)}

\providecommand{\idx}{\int_{\BR} \!\!\! dx}

\newcommand{\Ebb}{{\mathbb E}}

\providecommand{\eq}[1]{\begin{equation}\label{#1}}
\providecommand{\en}{\end{equation}}

\renewcommand{\d}{\displaystyle}

\providecommand{\qed}{\hfill {$\Box$}\par\medskip}
\providecommand{\BR}{{\mathbb{R}^3 }}
\providecommand{\BRN}{{\mathbb{R}^{3N} }}
\providecommand{\iijj}{\sum_{i,j=1}^N}
\renewcommand{\ij}{\sum_{i\not=j}^N}

\providecommand{\od}{{\rm OD}}
\providecommand{\dd}{{\rm D}}

\providecommand{\bi}
{\begin{itemize}}
\providecommand{\ei}{\end{itemize} }
\providecommand{\CC}{{\mathbb{C}}}
\providecommand{\RR}{\mathbb{R}}

\providecommand{\kak}[1]{(\ref{#1})}
\providecommand{\LR}{{L^2(\BR)}}

\providecommand{\LRN}{{L^2(\BRN)}}

\providecommand{\hni}{\hn_\ren }
\providecommand{\D}{{\ms D}}

\providecommand{\hn}{H}

\providecommand{\ms}[1]{\mathscr{#1}}

\providecommand{\ffff}{{\mathscr{F}_{\rm b}}}

\providecommand{\is}{\inf \s}

\providecommand{\lk}{\left(}
\providecommand{\rk}{\right)}
\providecommand{\lkk}{\left\{}
\providecommand{\rkk}{\right\}}

\providecommand{\add}{a^\ast}

\providecommand{\ass}{a^\sharp}

\providecommand{\ov}[1]{\overline{#1}}
\providecommand{\hf}{H_{\rm f}}

\providecommand{\gr}{\Psi_{\rm g}}
\providecommand{\JJ}{{\rm J}}

\providecommand{\grp}{\Psi_{\rm p}}
\providecommand{\half}{\frac{1}{2}}
\providecommand{\han}{{1/2}}

\providecommand{\hi}{H_{\rm I}}
\providecommand{\hp}{H_{\rm p}}

\providecommand{\la}{{\Lambda}}
\providecommand{\La}{{\Lambda}}

\providecommand{\s}{\sigma}
\providecommand{\hhh}{\ms H} 

\providecommand{\slim}{{\rm s}\!-\!\lim}

\providecommand{\vp}{{\widehat \varrho}}
\providecommand{\vpe}{{\widehat \varrho_\eps}}

\providecommand{\non}{\nonumber}

\providecommand{\IP}{{\rm P}}

\providecommand{\vp}{\mathop{\mathrm{{\varphi}}}\nolimits}

\providecommand{\e}{{\rm e}}

\makeatletter
\@addtoreset{equation}{section}
\makeatother

\providecommand {\R} {\ensuremath{\mathbb{R}}}

\providecommand {\F}  {\ffff}

\providecommand {\D}  {\ensuremath {\mathcal{D}}}
\providecommand {\s}  {\ensuremath {\mathcal{S}}}

\providecommand{\dd}{\mathop{\mathrm{d}}\nolimits}

\providecommand{\eps}{\varepsilon}

\providecommand{\ren}{{\rm ren}}
\providecommand{\ce}{{\rm c}_\eps}

\newcounter {constant}
\newenvironment{constant}{\refstepcounter{constant} }{}

\makeatletter
\@addtoreset{equation}{section}
\makeatother
\date{}
\begin{document}

\title
{{ Ultraviolet Renormalization of the Nelson Hamiltonian through
Functional Integration}}
\author{
\small Massimiliano Gubinelli\\[0.1cm]
 {\small\it    CEREMADE -- UMR 7534, Universit\'e Paris-Dauphine }    \\[-0.7ex]
  {\small\it  Place du Mar\'echal De Lattre De Tassigny,
75775 Paris cedex 16, France}      \\[-0.7ex]
 {\small  {\tt massimiliano.gubinelli@ceremade.dauphine.fr}   }\\[0.5cm]
\small Fumio Hiroshima\\
{\small\it Faculty of Mathematics, Kyushu University}    \\[-0.7ex]
{\small\it 744 Motooka, Fukuoka, 819-0395, Japan}      \\[-0.7ex]
{\small  {\tt hiroshima@math.kyushu-u.ac.jp}}\\[0.5cm]
\small J\'ozsef L\H{o}rinczi\\[0.1cm]
{\it \small School of Mathematics, Loughborough University} \\[-0.7ex]
{\it \small Loughborough LE11 3TU, United Kingdom} \\[-0.7ex]
{\small {\tt  J.Lorinczi@lboro.ac.uk}} \\[-0.7ex]}
\maketitle

\bigskip
\begin{abstract}
\noindent
Starting from the $N$-particle Nelson Hamiltonian defined by imposing an ultraviolet cutoff,
we perform ultraviolet 
renormalization by showing that in the zero cutoff limit a
self-adjoint operator exists after a logarithmically divergent term is subtracted from
the original Hamiltonian. We obtain this term as the diagonal part of a pair interaction
appearing in the density of a Gibbs measure derived from the Feynman-Kac representation
of the Hamiltonian. Also, we show existence of a weak coupling limit of the renormalized
Hamiltonian and derive an effective Yukawa interaction potential between the particles.

\end{abstract}

\newpage
\section{Introduction}
In this paper we consider the $N$-particle Nelson model, which describes the interaction
of $N$ electrically charged spinless particles with a scalar boson field.
In Fock space representation the model can be given by the Hamiltonian
\begin{equation} \label{Ham}
\hn = \hp\otimes \one +\one \otimes \hf  + \hi
\end{equation}
acting on
$$
\hhh=\LRN \otimes\ffff,
$$
where $L^{2}(\R^{3N})$ is the particle space and $\F$ denotes the symmetric Fock space over $\LR$
describing the bosons. Recall that $\F = \bigoplus_{n=0}^{\infty} \F^{(n)}$, where $\F^{(n)} =
{{\otimes}^n_{\rm sym}}\LR$ is the $n$-boson subspace and $\F^{(0)} = \mathbb{C}$ (where the
subscript indicates symmetrized tensor product), for which the infinite direct sum norm
$\|F\|_{\ffff }^2 = \sum_{n=0}^{\infty}\|f_n\|_{\F^{(n)}}^2$ is finite. We denote the Fock vacuum
by $\one_\ffff =1\oplus 0\oplus 0\oplus\ldots \in\ffff $, and write simply $\one$ when no confusion
arises.

The components of the $N$-particle Nelson Hamiltonian are as follows. The $N$-particle Schr\"odinger
operator
$$
\hp = -\frac{1}{2} \sum_{j=1}^N  \Delta_j + V
$$
is the Hamiltonian of the free particles with  an external potential $V: \BRN  \to \R$, which
acts as a multiplication operator, and where $\Delta_j=\Delta_{x_j}$ denotes the 3-dimensional
Laplacian.

Let $\add (f)$ and $a(f)$, $f\in\LR$, be the boson creation and annihilation operators, respectively,
satisfying the  canonical commutation relations
$$[a(f),\add(g)]=(\bar f, g),\quad
[a(f),a(g)]=0=
[\add(f), \add(g)].$$
We formally denote $\d \ass(f)=\int _\BR \ass(k)
f(k) dk$, where $\ass$
stands for $a$ or $\add$. Denote by $\omega(k)$ the dispersion relation, which we will choose in
the main part of the paper to be $\omega(k) = |k|$, describing massless bosons. The free field
Hamiltonian $\hf$ of $\ffff $ is then defined by the second quantization of $\omega$,
i.e.,
$$
\hf \prod_{j=1}^n \add(f_j)\one= \sum_{j=1}^n \add(f_1)\cdots\add(\omega f_j)\cdots\add(f_n)\one
\quad \mbox{and} \quad \hf \one=\one,
$$
formally expressed as
$$
\hf = \int_\BR \omega(k) \add(k)a(k)  dk.
$$

The interaction Hamiltonian is formally  defined by
\begin{equation}
\label{hint}
\hi(x)  = g\sum_{j=1}^N \int_\BR \frac{1}{\sqrt{2\omega(k)}}
\left( \vp (k)\e^{ik\cdot x_j}  a(k) + \vp (-k)\e^{-ik\cdot x_j} \add(k)\right)  dk
\end{equation}
for every $x\in\RR^{3N}$. We make the identification $\hhh\cong L^2(\RR^{3N};\ffff )$, i.e., $F\in \hhh$
will be regarded as a function  $\RR^{3N}\ni x\mapsto F(x)\in \ffff $ such that $\int_{\RR^{3N}}\|F(x)\|^2_
\ffff  dx<\infty$. Under this identification the interaction Hamiltonian becomes $(\hi F)(x)=\hi(x) F(x)$ on
$\hhh$. The function $\varphi$ (featured in its Fourier transform $\vp$) is a function describing a charge
distribution so that the total charge is $\int_\BR \varphi(x) dx = 1$. In the Hamiltonian it has a
regularising role, making the operator well defined, and physically it has the meaning of an ultraviolet
(UV) cutoff. The prefactor $g$ is a coupling constant between the particles and the field.

Under the assumptions
\begin{equation}
{\vp }/{\omega^{1/2}}, {\vp }/{\omega}  \in   \LR ,\quad \ov{\vp(k)}=\vp(-k)
\label{sa}
\end{equation}
the interaction $\hi$ is well defined, symmetric and infinitesimally $\one\otimes \hf $-bounded. Thus
by the Kato-Rellich theorem $\hn$ is
 self-adjoint   $D(\hp\otimes \one) \cap
D(\one \otimes \hf )$. If, moreover, an infrared (IR) cutoff is imposed by the condition
\begin{equation}
{\vp }/{\omega^{3/2}}  \in  \LR,
\label{ir}
\end{equation}
then $\hn$ has a unique ground state \cite{spo98,bfs98,ger00,ara01,sas05}, i.e., an eigenfunction $\Psi
\in \hhh$ at the bottom of its spectrum. As it was shown in \cite{lms02,hir06}, condition (\ref{ir})
is also necessary for a ground state to exist in this space.

In this paper we are interested in an appropriate definition of $\hn$ in the point charge limit, i.e., when
$\varphi(x) \to (2\pi)^{3/2}\delta(x)$ or, equivalently, $\vp (k) \to 1 $. This is a physically interesting
but mathematically singular case, when condition (\ref{sa}) fails to hold. In order to analyse this limit,
we regularise the Hamiltonian by choosing the UV cutoff function $\vpe(k)=\e^{-\eps|k|^2/2}$. With this
choice in (\ref{hint}) we define the Hamiltonian $\hn_\eps$, and by regarding $\eps>0$ as a UV cutoff parameter
we will analyse the limit of $\hn_\eps-E_\eps$ as $\eps\downarrow 0$, where $E_\eps$ is an energy renormalization
term, which will be determined below.

The main results of this paper are as follows.
\bi
\item[(1)]
By using functional integration we derive the energy renormalization term $E_\eps$ from the diagonal part of
a pair interaction, and show the existence of the renormalized Hamiltonian $\d \hn_\ren =\lim_{\eps\downarrow 0}
(\hn_\eps-E_\eps)$ in the sense of strong convergence of the related semigroups.
\item[(2)]
We derive the pair interaction potential in the 
path measure associated with
$\hni$.
\item[(3)]
We show existence of the weak coupling limit  of $\e^{-t\hni}$ and compare it with that of the Nelson model
with UV cutoff.
\ei
Here are some comments to these points.

\medskip
\noindent
(1) The first part of this problem was already investigated in \cite{nel64} by using Gross transform. This
is a unitary transform implemented by
$e^{i\pi_\eps(x)}$ with the generator
\eq{gross}
\pi_\eps(x)=i\sum_{j=1}^N \int_{|k|>\Lambda} \frac{1}{\sqrt{2\omega(k)}}\frac{1}{\omega(k)+|k|^2/2}
(-a(k)\vpe(k)e^{ikx_j}+\add(k)
\vpe(-k)e^{-ikx_j})dk.
\en
Here $\La>0$ is introduced  for $\pi_\eps$ to be well defined. Nelson has shown that the Gross transformed
operator $\e^{i\pi_\eps(x)} (H_\eps-E_\eps) \e^{-i\pi_\eps(x)}$ converges to
a self-adjoint operator in norm
resolvent sense as $\eps\downarrow 0$, and $\e^{i\pi_\eps(x)}\to \e^{i\pi_0(x)}$ in strong sense.

In contrast to this approach, in the following we present a UV renormalization by using path measure methods.
Nelson did consider a path integral approach \cite{nel64b}, however, this remained incomplete since the approach
based on Gross transform may have appeared simpler and satisfactory for the purposes of his investigation.
Taking the Gross transformation of $\hn_\eps-E_\eps$, a cancellation of diverging terms occurs and the limit
$\eps\downarrow 0$ can be analysed to define a UV renormalised Hamiltonian. However, in this paper we do not
take Gross transform and derive $E_\eps$ from the diagonal part of the pair potential associated with a Gibbs
measure instead.

Following our previous work on 
the Nelson model \cite{lms02,bhlms02} we find it worthwhile to
analyse this problem by using functional integration not just for the extra insights it gives (applicable
also to other cases, e.g., UV renormalization of the Nelson model with variable coefficients \cite{ghps12}),
but also because this approach allows to prove existence of a ground state of the UV renormalized Hamiltonian.
As far as we are aware, the existence of a ground state of $H_\ren$ was shown for sufficiently weak couplings
only \cite{hhs05}. This problem is another illustration of the fact that direct analytic and path integral
methods complement each other, and both have specific advantages.

A key point in this paper is to show that
\eq{DD}
\lim_{\eps\downarrow0}
(f\otimes\one, \e^{-T(H_\eps-E_\eps)}g\otimes \one)
\en
can be expressed in terms of a path measure representation
(Lemma \ref{maindenai} below), and
\eq{CC}
H_\eps-E_\eps>C
\en
holds with a constant $C$, uniformly in $\eps>0$ (Corollary \ref{nelson}). Although these were established by
operator analysis methods in \cite{nel64} by using the Gross transform, we derive them directly by using
Feynman-Kac type formulae for $H_\eps-E_\eps$, so our strategy is completely different from Nelson's.

\medskip
\noindent
(2)
By constructing a functional integral representation of the Nelson Hamiltonian with UV cutoff and integrating out
the field part, an expression is obtained in terms of an expectation with respect to Wiener measure on Brownian
paths under $V$ and a pair interaction potential $W$ (see Chapter 6 in \cite{lhb11}). Using this as a density,
the path measure can be seen as a Gibbs measure on paths. The pair interaction has the form
\eq{p1}
\int_{-T}^ T ds \int_{-T}^ T dt W(B_t-B_s,t-s),
\en
where $W$ depends on the UV cutoff \cite{bhlms02}. On the other hand, a similar construction for the Pauli-Fierz
model in non-relativistic quantum electrodynamics yields a Gibbs measure with pair interaction formally given by
\cite{spo87,bh09}
\eq{p2}
\int_{-T}^ T dB_s^\mu  \int_{-T}^ T dB_t^\nu  W_{\mu\nu}(B_t-B_s,t-s).
\en
It is remarkable that the double Riemann integral in \kak{p1} is replaced by a double stochastic integral.
In this paper we will consider the finite volume Gibbs measure associated with the Nelson model without UV
cutoff and obtain that the exponent in the Boltzmann-Gibbs density has the form
(Corollary \ref{pair potential})
\eq{p3}
\int_{-T}^ T ds \int_{s}^ T dB_t  W(B_t-B_s,t-s) +\int _{-T}^T ds W(B_T-B_s,T-s).
\en
It is interesting to see that the Gibbs measure without a UV cutoff has the intermediate form of \kak{p1} and
\kak{p2}. The representation of the renormalized pair potential in terms a mixed integral is obtained via an 
It\^o formula. This technique has been widely used to study the intersection local time of Brownian 
motion~\cite{Yo1,Yo2,Yo3} and can be used to study the related polymer measure in two dimensions ~\cite{LeG1,LeG3}. 
A further application of the It\^o formula could in principle transform the pair potential given by a double 
Lebesgue integral \kak{p1} into a pair potential given by a double It\^o integral similar to~\kak{p2}. However, 
as analyzed in some dept in \cite{FG1} in the context of a stochastic model for 3d vortex filaments, pair 
potentials given by double stochastic integrals are difficult to  handle analytically and do not allow for uniform 
exponential estimates (see also \cite{GL}). The mixed representation we have chosen is better suited for bounds 
which are valid for any value of the coupling constant $g$.

\medskip
\noindent
(3)
Finally we consider the weak coupling limit, which  is a scaling limit such that the creation operators $\add$
and the annihilation operators $a$ are scaled to $\kappa a$ and $\kappa \add$, respectively, with a scaling
parameter $\kappa>0$. When $\eps>0$, the scaled Nelson Hamiltonian $\hn_\eps(\kappa)$ converges in the limit
$\kappa \to \infty$ to a Schr\"odinger operator with an effective potential, and furthermore it converges to
a Schr\"odinger operator with Yukawa potential when in a subsequent limit $\eps\downarrow 0$. We are interested
in obtaining the weak coupling limit of $\hni(\kappa)$ and comparing it with that of the UV-regularized
Hamiltonian (Corollary \ref{main5}).

\bigskip
We note that in this paper the space dimension is fixed to the physical $d=3$, however, our arguments can
be carried out for any other dimension in a similar way. We will see that the energy renormalization term
behaves like $E_\eps\sim -\int_1^\infty e^{-\eps r^2} r^{d-4} dr$. When $d=1$ or $d=2$, there is no need
for an energy renormalization, and when $d\geq 3$, energy renormalization becomes important to deal with
the UV divergences in the $\eps\downarrow 0$ limit. Also, it will be seen that our arguments do not need
the assumption of a pinning external potential, in fact, we do not need any. However, we keep a $V$ in our
formulae with the understanding that the results are valid also for $V \equiv 0$.

\bigskip
\noindent
The plan of this paper is as follows. In the main Section 2 we perform renormalization on the level of the
density of the Gibbs measure, determine the pair interaction potential, and prove existence of a UV
renormalized Hamilton operator. In Section 3 we study a weak coupling limit of the renormalized Hamiltonian
and derive an effective interaction potential between the particles.

\section{Energy renormalization by path measures}
\subsection{Functional integral representation of  regularized Hamiltonians}
First we define the version of the Nelson model which will be the main object studied in this paper. We choose
$\omega(k)=|k|$. Let
$$
\one_\la(k) =
\lkk
\begin{array}{ll}
1,& \omega(k) < \la \\
0,& \omega(k)\geq\la
\end{array}
\right.
$$
and $\kp=\one-\one_\la(k)$. We assume that  $\Lambda> 0$, which is needed in \kak{IRbound},  Lemma
\ref{finite} and Corollary \ref{nelson} below. For simplicity we will use the following standing assumption
throughout below.
\begin{assumption}
The external potential $V$ is a bounded continuous function. In particular, it is of Kato-class, i.e.,
it satisfies
\eq{kato11}
\lim_{t\downarrow 0}\sup_{x\in\RR^3} \Ebb^x\!\left [\int_0^t |V(B_s)|ds\right ]=0.
\en
\end{assumption}
Note that in what follows $V\equiv 0$ is a possible choice without changing the statements. We summarize
some properties of Kato-class potentials in Appendix \ref{appa} for the reader's convenience; they will be
used below on some objects which take the role of a potential.

Consider the cutoff function
\eq{sasa3}
\vpe(k)=\e^{-\eps|k|^2/2}\kp,\quad \eps\geq0
\en
and define the regularized Hamiltonian
\begin{equation}
\hn_\eps = \hp\otimes \one +\one \otimes \hf  + \hi^\eps,\quad \eps>0,
\end{equation}
by
\begin{equation}
\hi^\eps(x)  = g\sum_{j=1}^N \int_\BR
\frac{1}{\sqrt{2\omega(k)}} \left( \vpe (k)\e^{ik\cdot x_j}
 a(k) + \vpe (-k)\e^{-ik\cdot x_j}  \add(k)\right)  dk.
\end{equation}
Here $\eps>0$ is regarded as the UV cutoff parameter. The main purpose of this paper is to
consider the limit $\eps\downarrow 0$ of $\hn_\eps$. We show that this limit can be sensibly defined
by an energy renormalization. Define
\eq{sasa6}
E_\eps=-\frac{g^2}{2} N \int_{\BR} \frac{\e^{-\eps|k|^2}}{\omega(k)}\beta(k)\kp dk,
\en
where
\eq{beta}
\beta(k)=\frac{1}{\omega(k)+|k|^2/2}.
\en
Notice  that $E_\eps\to-\infty$ as $\eps\downarrow 0$.

Our main theorem states that $\hn_\eps-E_\eps$  converges in the $\eps\downarrow 0$ limit to a
non-trivial self-adjoint operator $\hni$ in a specific sense, which we call the UV renormalized
Nelson Hamiltonian.
\bt{main}
There exists a self-adjoint operator $\hni$ such that
\eq{main1}
\slim_{\eps\downarrow0}
\e^{-t(\hn_\eps-E_\eps)}=\e^{-t\hni},\quad t\geq0.
\en
\et
We carry out the proof by functional integration and will obtain $E_\eps$ as the diagonal term of a
pair interaction potential on the paths of a Brownian motion.

In the following we will fix some time interval $[-T,T]$ once for all and track the dependence on $T$ of the various estimates.
A Feynman-Kac formula holds
for $(F,\e^{-2T\hn_\eps}G)$ (see \cite[Theorem  6.3]{lhb11}). In particular, for $F=f\otimes \one$ and $G=h\otimes \one$ it can be described in terms of a family of $N$ indepedent two-sided $\BR$-valued Brownian motions $(B_t)_{t\in\RR}=(B_t^1,...,B_t^N)_{t\in\RR}$, where $(B_t^j)_{t\in\RR}$, $j=1,...,N$, are $N$
independent, two-sided, $\BR$-valued standard Brownian motions. It is convenient to take  $(B_t)_{t\in\RR}$ to be the canonical process on the space of $\BRN$-valued continuous paths indexed
by the whole real line, endowed with Wiener measure $\IP^x$ starting from  $x\in (\BR)^N$ at $t=-T$.  We will denote by $\Ebb^x$ the associated expectation. Note that with respect to $\IP^x$ the process $(B_t)_{t\ge -T}$ is a martingale with respect to the forward
filtration
\eq{filtration}
{\cal F}^T=({\cal F}_t^T)_{t\ge -T}
\en
 with ${\cal F}_t^T=\sigma(B_s: -T \le s\le t)$.

\begin{proposition}
Let $f,h\in L^2(\BRN)$ then
\begin{align}
\label{eq:fk-base}
(f\otimes \one, \e^{-2T\hn_\eps} h\otimes  \one) = \int_{\BRN} dx\, \Ebb^x\!\left[\ov{f(B_{-T})} h(B_T)
\e^{-\int_{-T}^T V(B_s) ds} \e^{\frac{g^2}{2} S_\eps^T} \right],
\end{align}where
\eq{regint}
S_\eps^T= \iijj \int _{-T}^T ds\int_{-T}^T dt
\, W_\eps (B_t^i-B_s^j,t-s)
\en
is the pair interaction given by the pair interaction potential $W_\eps: \BR\times \RR \to \RR$ given by
\begin{equation}
\label{eq:nelsonWreg}
W_\eps (x,t) =  \int_{\BR} \frac{1}{2\omega(k)} \e^{-\eps |k|^2} \e^{-ik\cdot x} \e^{-\omega(k)|t|}\kp dk.
\end{equation}
\end{proposition}

\subsection{Renormalized action}
In this section we perform UV renormalization on the level of the density of the path measure~\eqref{eq:fk-base}.
 Consider the
function
\begin{equation}
\label{eq:phiW}
\varphi_\eps(x,t) =  \int_{\BR} \frac{\e^{-\eps |k|^2} \e^{-ik\cdot x-\omega(k)|t|}}{2 \omega(k)}
\beta(k)\kp dk,\quad \eps\geq 0,
\end{equation}
where $\beta(k)$ is given by \kak{beta}.

\begin{proposition}
There exists a functional $S_0^\ren $ such that
\eq{mmm1}
\lim_{\eps\downarrow 0}\Ebb^x \left[\e^{-\int_{-T}^TV(B_s)ds} \e^{\frac{g^2}{2}( S_\eps-4NT\varphi_\eps(0,0))}\right] =
\Ebb^x\!\left[\e^{-\int_{-T}^TV(B_s)ds} \e^{\frac{g^2}{2}S_0^\ren }\right].
\en
\end{proposition}
\noindent
Notice that $W_\eps (x,t)$ is smooth, and
$W_\eps (x,t)\to  W_0(x,t)$ as $\eps\downarrow0$ for every
$(x,t)\neq (0,0)$, where
\eq{tooi}
W_0(x,t)=\int_{\BR}\frac{1}{2\omega(k)}\e^{-ik\cdot x} \e^{-\omega(k)|t|} \kp dk.
\en
It is seen, however, that $W_\eps (0,0)\to \infty$ as $\eps\downarrow 0$, i.e., $W_0(x,t)$ has a power-like
singularity at $(0,0)$, thus \kak{mmm1} is non-trivial to obtain. We will prove the above proposition
through a sequence of lemmas below.

 As $\eps\downarrow 0$ only the diagonal part of the interaction has a singular term.
Fix $0<\tau\leq T$ and denote $[t]_T = -T\vee t \wedge T$. We decompose the regularized interaction
into diagonal and off-diagonal parts as
\eq{hiror6}
S_\eps^T = S_\eps^{\dd,T } + S_\eps^{\od,T },
\en
where
\eq{eniwa1}
S_\eps^{\dd,T } = 2\iijj \int_{-T}^T ds \int_{s}^{[s+\tau]_T} dt\, W_\eps (B_t^i-B_s^j,t-s)
\en
and
\eq{eniwa2}
S_\eps^{\od,T }  = 2\iijj \int_{-T}^T ds \int_{[s+\tau]_T}^{T}dt W_\eps (B_t^i-B_s^j,t-s).
\en
$S_\eps^{\dd }$ denotes the integral of $S_\eps$ around the diagonal $\{(t,t)\in\RR^2| |t|\leq T\}$, 
and $S_\eps^{\od } $ in its complement. Notice that for $\tau=T$ we have $S_\eps^{\od }=0$. The next 
lemma is easy to prove and we omit the details.
\bl{h1}
We have  that $\lim_{\eps\downarrow 0}S_\eps^{\od,T }  = S_0^{\od,T }$ pathwise.
\el

A representation in terms of a stochastic integral will help us deal with the more difficult term
$S_\eps^{\dd,T }$. In the following lemma first we derive some bounds on $\varphi_\eps(x,t)$ and its
gradient.
\begin{lemma}
\label{lpb}
There exists a constant $c>0$ such that the bounds
\begin{align*}
&
|\nabla \varphi_\eps(x,t)|\le c  |t|^{-1},\quad t\not=0  \\
&
|\nabla \varphi_\eps(x,t)| \le c |x|^{-1},\quad |x|\not=0
\end{align*}
hold uniformly in $\eps$. Furthermore, similar bounds hold for the function $\varphi_0-\varphi_\eps$
with a constant $\ce   > 0$ such that $\lim_{\eps\downarrow 0}\ce  =0$, i.e.,
\begin{align*}
&
|\nabla \varphi_\eps(x,t)-\nabla \varphi_0(x,t) |\le \ce    |t|^{-1},\quad t\not=0,  \\
&
|\nabla \varphi_\eps(x,t)-\nabla \varphi_0(x,t) | \le \ce   |x|^{-1},\quad |x|\not=0.
\end{align*}\end{lemma}
\proof
The first bound on the gradient follows directly by
\begin{equation*}
\begin{split}
|\nabla \varphi_\eps(x,t)| & \le
\int_{\BR} \frac{1}{2 (\omega(k)+|k|^2/2)} \e^{-\eps |k|^2} \e^{-\omega(k)|t|}\kp dk
\le c \int_\La^\infty  \e^{ -r t} dr.
\end{split}
\end{equation*}
Next consider the second. After integration over the angular variables we obtain
\eq{eq:varphi-ang-int}
\varphi_\eps(x,t) = 2\pi \int_{\Lambda}^\infty \frac{\e^{-\eps r^2 - r |t|}}{r(2+r)} \frac{\sin(r|x|)}{|x|} dr.
\en
Differentiation in (\ref{eq:varphi-ang-int}) gives
\begin{equation}
\label{b}
\nabla \varphi_\eps(x,t)
= \frac{2\pi x}{|x|^2}
\int_{\Lambda|x|}^\infty  \frac{\e^{-\eps r^2/|x|^{2}-|t| r/|x|}}{r(2|x|+r)}
\lk r\cos r -\sin r \rk dr,
\end{equation}
and estimating the right-hand  side we have
\begin{align*}
|\nabla \varphi_\eps(x,t)|
 \leq \frac{1}{|x|}
 \lk
 \int_{0}^{1}  \frac{C r^3}{r^2} dr
+ \left|\int_{1}^\infty   \frac{\e^{-\eps r^2/|x|^{2}-|t|r/|x|}}{(2|x|+r)} \cos r dr\right|
+ \int_{1}^\infty  \frac{1}{r^2}dr\rk
\end{align*}using that $|r\cos r  - \sin r | \le C r^3$, for a constant $C > 0$ and all $r\in [0,1]$. Since the
integral in the middle term above is bounded, the lemma follows.
\qed

Define
\begin{align}
\label{eq:WI-renorm-1}
&X_\eps^T=2\ij  \int_{-T}^{T}\varphi_\eps(B_s^i-B_s^j,0) ds,\\
&Y_\eps^T=2\sum_{i,j=1}^N  \int_{-T}^{T} ds \int_s^{[s+\tau]_T}\nabla\varphi_\eps(B_t^i-B_s^j, t-s) \cdot dB_t^i,\\
\label{eq:WI-renorm-2}
&Z_\eps^T=-2\sum_{i,j=1}^N \int_{-T}^{T}\varphi_\eps(B_{[s+\tau]_T}^i-B_s^j, [s+\tau]_T-s)ds.
\end{align}

\begin{lemma}
\label{lem:WI-decomp}
The representation formula
\begin{align}
S_\eps^{\dd,T } & = 4TN \varphi_\eps(0,0) + X_\eps^T + Y_\eps^T  + Z_\eps^T
\label{eq:WI-decomp}
\end{align}
holds for all $\eps>0$.
\end{lemma}
\proof
Note that $\varphi_\eps(x,t)$ solves the equation
\begin{equation}
\label{eq:phi-pde}
\lk \partial_t  + \frac12 \Delta\rk  \varphi_\eps(x,t) = - W_\eps (x,t), \qquad x \in \BR, \quad t \ge 0.
\end{equation}
Fix $i$ and $j$. The  It\^o formula yields that
\begin{multline}
\label{eq:phi-eps-dec}
\varphi_\eps(B_{[s+\tau]_T}^i-B_s^j,[s+\tau]_T-s)-\varphi_\eps(B_s^i-B_s^j,0) \\
= \int_{s}^{[s+\tau]_T} \nabla \varphi_\eps(B_t^i-B_s^j,t-s)  \cdot dB_t^i
+ \int_{s}^{[s+\tau]_T}
\lk \partial_t + \frac{1}{2}\Delta\rk  \varphi_\eps(B_t^i-B_s^j,t-s) dt.
\end{multline}
Hence by (\ref{eq:phi-pde})
\begin{multline}
\label{eq:Wdec}
\int_{s}^{[s+\tau]_T} W_\eps (B_t^i-B_s^j,t-s) dt \\
= \varphi_\eps(B_s^i-B_s^j,0) -\varphi_\eps(B_{[s+\tau]_T}^i-B_s^j,[s+\tau]_T-s) +
\int_{s}^{[s+\tau]_T} \nabla\varphi_\eps(B_t^i-B_s^j,t-s)\cdot dB_t^i
\end{multline}
follows. Inserting this expression into $S_\eps^{\dd,T}$ proves the claim.
\qed
Lemma~\ref{lem:WI-decomp} suggests the definition
\eq{jozsef def}
S_\eps^\ren  = S_\eps^T - 4 NT\varphi_\eps(0,0),
\en
as a renormalized action.
This can be expressed as
\begin{align}
\label{eq:WI-renorm}
S_\eps^\ren
=
S_\eps^{\od,T }  +X_\eps^T +Y_\eps^T +Z_\eps^T.
\end{align}

\bl{finitez}
Let $\eps\geq0$.
There exist constants $c_Z$ and $c_S$ such that we have $|S_\eps^{\od,T}|\leq c_S (T+1)$
and $|Z_\eps^T|\leq c_Z T$, uniformly in the paths and in $\eps\geq0$.
\el
\proof
We see that
\begin{align*}
|Z^T_\eps| \leq
4\pi N^2 \lk \int_{-T}^{T-\tau} ds  \int_{\Lambda}^\infty \frac{\e^{-r\tau }}{1+r/2}dr +
\int_{T-\tau}^{T} ds  \int_{\Lambda}^\infty \frac{\e^{-r(T-s)}}{1+r/2}dr\rk \leq c_z T
\end{align*}
with some $c_z > 0$. It is also easy to see that $|S_\eps^{\od,T}|\leq {\rm const} \lk 
(\frac{2T}{\tau}-1)+\log(\frac{\tau}{2T})\rk$, hence the bound $|S_\eps^{\od,T}|\leq c_s (T+1)$ 
follows.
\qed
\bl{finite}
Let $\eps\geq0$. There exists a  constant $c_X$ independent of $\eps$ such that for all 
$\alpha >0$ and $T>0$ we have
$$
\d \sup_{x\in\BRN} \Ebb^x[e^{\alpha|X^T_\eps|}]\le e^{\alpha c_X T}.
$$
 \el
\proof
We notice that
\eq{225}
X_\eps^T =\ij\int_{-T}^ T ds \frac{2\pi}{|B_s^i-B_s^j|}
\int_\Lambda^\infty \frac{\sin\sqrt{r|B_s^i-B_s^j|}}{r+r^2/2}\e^{-\eps r^2} dr,\quad \eps\geq0.
\en
The assumption $\Lambda>0$ implies
\eq{IRbound}
a=2 \pi \int_\Lambda^\infty \frac{1}{r+r^2/2} dr<\infty.
\en
Hence
\eq{xkato}
|X^T_\eps| \leq a \ij\int_{-T}^ {T}  \frac{ds}{|B_s^i-B_s^j|}
\en
Since $\ij |x^i-x^j|^{-1}$  is a Kato-class potential on $L^2(\RR^{3N})$ (see Appendix \ref{appa}),
the claim follows.
\qed
By the stochastic Fubini theorem we can interchange the stochastic and Lebesgue integrals when $\eps>0$. 
Thus $Y^T_\eps$ has the representation
\begin{equation}
\label{eq:WI-renorm-1-bis}
Y_\eps^T  = \sum_{i=1}^N \int_{-T}^T \Phi^i_{\eps,t} dB_t^i,\quad \eps>0,
\end{equation}
where $\Phi_{\eps,t}=(\Phi^1_{\eps,t},\dots,\Phi^N_{\eps,t})$ is the process with values in $\BRN$ 
given by
$$
\Phi^i_{\eps,t} = 2 \sum_{j=1}^N \int_{[t-\tau]_T}^t  \nabla \varphi_\eps(B_t^i-B_s^j,t-s) ds.
$$
Define
\begin{equation}
\label{eq:WI-renorm-1-biss}
Y_0^T  = \sum_{i=1}^N \int_{-T}^T \Phi^i_{\eps,0} dB_t^i.
\end{equation}
\begin{lemma}
\label{prop:convergence}
Let $\eps\geq0$. Then there exists a  constant $c_Y$ independent of $\eps$ such that for all $\alpha >0$
it follows that 
$$
\d \sup_{x\in\BRN} \d \Ebb^x [e^{\alpha Y^T_\eps}]\le e^{c_Y\alpha^2 T} \quad \mbox{and} \quad
\d \lim_{\eps\downarrow0}\Ebb^x [|Y^T_\eps -Y_0^T |^2]=0
$$
for all $x\in\BRN$.
\end{lemma}
\proof
The process $(\Phi^i_{\eps,t})_{t\in\RR}$ is adapted to the forward filtration ${\cal F}^T$ given by 
\kak{filtration} so $Y_\eps^T$  is the terminal value of a martingale with quadratic variation given 
by the $L^2([-T,T];\BRN)$ norm of $\Phi_{\eps,\cdot}$. We have
\begin{eqnarray*}
\int_{-T}^{T} |\Phi_{\eps,t}|^2 dt  
&\leq& 
4 \sum_{i=1}^N \int_{-T}^{T} \left[ \sum_{j=1}^N \int_{[t-\tau]_T}^t  
|\nabla \varphi_\eps(B_t^i-B_s^j,t-s)| ds \right]^2 dt\\
&\leq& 
4c^2  N\iijj\int_{-T}^{T} 
\left[ \int_{[t-\tau]_T}^t  |B_t^i-B_s^j|^{-\theta}|t-s|^{-(1-\theta)} ds \right]^2 dt,
\end{eqnarray*}
where we used Jensen's inequality, Lemma \ref{lpb} and an interpolation to obtain the bound
\eq{interp1}
|\nabla \varphi_\eps(x,t)| \le c   |x|^{-\theta} |t|^{-(1-\theta)}, \quad \theta \in [0,1],
\en
which is uniform in $\eps\in[0,1]$. With some $\frac{1}{2}<\theta <1$, the Cauchy-Schwarz inequality applied
to the latter integral gives
\begin{align*}
\int_{-T}^{T} |\Phi_{\eps,t}|^2 dt
&\leq 4c^2  N\iijj  \int_{-T}^T \left[\int_{[t-\tau]_T}^t |B_t^i-B_s^j|^{-2\theta} ds \right]
\lk \int_{[t-\tau]_T}^t \!\!\! |t-s|^{-2(1-\theta)} ds \rk  dt\\
&
\leq  4c^2  \tau^{2\theta-1}  N\iijj \int_{-T}^T \left[\int_{[t-\tau]_T}^t |B_t^i-B_s^j|^{-2\theta} ds \right]  dt\\
&
\leq 4 c^2 \tau^{2\theta-1} N  Q,
\end{align*}
where $c$ is the constant in Lemma \ref{lpb}, which is independent of $\eps$ and
$$
Q= \iijj \int_{-T}^{T} ds \int_s^{[s+\tau]_T} |B_t^i-B_s^j |^{-2\theta} dt.
$$
Then, by the Girsanov theorem,
\begin{align}
\lk \Ebb^x\!\left[\e^{\alpha Y_\eps^T}\right] \rk ^2
&\leq
\Ebb^x\!\left[\e^{2\alpha \int_{-T}^ {T} \Phi_{\eps,t} \cdot dB_t-\half(2\alpha )^2\int_{-T}^ {T} |\Phi_{\eps,t}|^2 dt}\right]
\Ebb^x\!\left[\e^{2\alpha ^2 \int_{-T}^ {T} |\Phi_{\eps,t}|^2 dt}\right] \nonumber\\
&=
\Ebb^x\!\left[\e^{2\alpha ^2\int_{-T}^{ T} |\Phi_{\eps,t}|^2 dt}\right]
\leq
\Ebb^x\!\left[\e^{\gamma  Q}\right].
\label{na2-occ1}
\end{align}
where  $\gamma=8 c\sqrt N \alpha^2\tau^{2\theta-1}$ and where we recall that we have chosen $\half <\theta <1$.
Jensen's inequality gives
\eq{domo}
\Ebb^x\!\left[\e^{\gamma Q} \right]  \le \int_{-T}^{T} \frac{ds}{2T}
\Ebb^x\!\left[\e^{2 T\gamma \iijj \int_s^{[s+\tau]_T} |B_t^i -B_s^j|^{-2\theta} dt} \right].
\en
Note that
$[s+\tau]_T \le s+\tau$ so the right hand side
 is bounded by
\eq{domo-2}
\Ebb^x\!\left[\e^{\gamma Q} \right]  \le \int_{-T}^{T} \frac{ds}{2T}
\Ebb^x\!\left[\e^{2 T\gamma \iijj \int_s^{s+\tau} |B_t^i -B_s^j|^{-2\theta} dt} \right].
\en
Taking conditional expectation with respect to
$({\cal F}_s)_{s\geq0}$ with
${\cal F}_t=\s(B_r, 0\leq t)$, and using the Markov property, we see that
\begin{align*}
\Ebb^x \left[\e^{2 T\gamma \iijj \int_0^{\tau} |B_{s+t}^i-B_s^j|^{-2\theta} dt }\right]
&=
\Ebb^x\!\left[
\Ebb^x \left[\e^{2 T\gamma \iijj \int_0^{\tau} |B_{s+t}^i-B_s^j|^{-2\theta} dt }|{\cal F}_s\right]\right]\\
&=
\Ebb^x\!\left[\Ebb^{B_s} \left[\e^{2 T\gamma \iijj \int_0^{\tau}
|B_{t}^i-B_0^j|^{-2\theta} dt }\right]\right].
\end{align*}
Since $|x|^{-2\theta}$ is a Kato-class potential, we see that
\eq{tau}
\d \sup_{x,z\in\BR}\Ebb^x [\e^{\beta \int_0^\tau |B^i_s+z|^{-2\theta}ds}]=
\sup_{x\in\BR} \Ebb^x [\e^{\beta \int_0^\tau |B^i_s|^{-2\theta}ds}] \le e^{c \tau\beta},
\en
for some $c>0$ and all $\beta>0$ (see Appendix \ref{appa}). From this it follows that
\eq{suppp}
\sup_{x\in\BRN} \Ebb^x\!\left[\e^{\gamma Q} \right] \le \sup_{x\in\BRN} \int_{-T}^{T}
\frac{ds}{2T} \Ebb^x \left[\e^{2 T\gamma \iijj \int_0^{\tau} |B_{s+t}^i-B_s^j|^{-2\theta} dt }\right]
\le  e^{\alpha^2 cT}
\en
for possibly a different constant $c$ (here and in the following formula). Hence we obtain
\eq{sup3}
\sup_{\eps\in(0,1]}\sup_{x\in\BR}\Ebb^x[\e^{2\alpha Y_\eps^T}] \leq
e^{ \alpha^2 c T}.
\en
for all $\alpha\in \RR$. By similar computations we can establish that the process $\Phi_\eps$ converges 
to $\Phi_0$ in $L^2([-T,T],\BRN)$ almost surely under Wiener measure $\IP^x$, for all $x\in\BRN$. For every 
$0<\eps$ we have indeed
\begin{align*}
&\int_{-T}^{T} |\Phi_{\eps,t}-\Phi_{0,t}|^2 dt  \\
&\leq 4\ce^2  N\iijj  \int_{-T}^T \left[\int_{[t-\tau]_T}^t |B_t^i-B_s^j|^{-2\theta} ds \right]
\lk \int_{[t-\tau]_T}^t \!\!\! |t-s|^{-2(1-\theta)} ds \rk  dt\\
&\leq  4\ce^2  \tau^{2\theta-1}  N\iijj \int_{-T}^T \left[\int_{[t-\tau]_T}^t |B_t^i-B_s^j|^{-2\theta} ds \right]  dt\\
& \leq 4 \ce^2 \tau^{2\theta-1} N  Q,
\end{align*}
where as above we used  Lemma \ref{lpb} and an interpolation to obtain the bound
\eq{interp2}
|\nabla \varphi_\eps(x,t)-\nabla \varphi_{0}(x,t)| \le \ce    |x|^{-\theta} |t|^{-(1-\theta)},
\quad \theta \in [0,1]
\en
for all $0<\eps$ and where the constant $\ce$ is such that $\ce  \to 0$ as $\eps\downarrow 0$. The almost 
sure convergence is then a consequence of the fact that we have already shown that $Q<+\infty$ almost 
surely under $\IP^x$ for all $x\in\BRN$. The convergence of $\Phi_\eps$ implies also the convergence of the 
martingale $Y^T_\eps$ to $Y^T_0$, at least in $L^2(\Omega,\IP^x)$ for all $x\in\BRN$.

\qed

\begin{lemma}
\label{prop:exp-int}
There exists a constant $c_{ren}$ such that for all $\alpha\in\RR$ and  $f,h\in L^2(\BRN)$ we have
$$
\d \int_{\BRN} dx \Ebb^x[f(B_{-T}) h(B_{T}) \e^{-\int_{-T}^TV(B_s) ds}
\e^{\alpha S_\eps^\ren }]\le  \|f\| \|h \| e^{c_\ren(\alpha^2 T+\alpha T + \alpha )}
$$
for all $\eps\geq0$.
\end{lemma}
\proof
Recall the decomposition $S_\eps^\ren  = S_\eps^{\od}+X_\eps+Y_\eps + Z_\eps$. By the Cauchy-Schwarz
inequality we have
\begin{align}
&
\int_{\BRN} dx \Ebb^x[|f(B_{-T}) h(B_T)| \e^{\alpha S_\eps^\ren }]\non\\
&=
\int_{\BRN} dx \Ebb^0\!\left[|f(x) h(B_{T})|\e^{-\int_{-T}^{T}V(B_s) ds}
\e^{\alpha(
S_\eps^{\od, T}+X_\eps^T+Y_\eps^T+Z_\eps^T)
} \right]\non\\
&\label{tosan}
\leq \|f\| \|h \| \sup_{x\in\BRN}
\lk
\Ebb^x
\left[\e^{-2\int_{-T}^{T}V(B_s) ds} \e^{2\alpha(
S_\eps^{\od, T}+X_\eps^T+Y_\eps^T+Z_\eps^T)}
\right]\rk^\han.
\end{align}
By Lemmas \ref{finitez}, \ref{finite} and \ref{prop:convergence} and the fact that $V$ is Kato-class, we see that there exists a constant $c_{\ren}$ such that
$$
\sup_{x\in\BRN} \Ebb^x\left[\e^{-2\int_{-T}^{T}V(B_s) ds} \e^{
2\alpha(
S_\eps^{\od, T}+
X_\eps^T+Y_\eps^T+Z_\eps^T)}\right]\le e^{2 c_\ren(\alpha^2 T+\alpha T+\alpha )},
$$
and the lemma follows.
\qed

\subsection{Renormalized Hamiltonian}
In this section we show that $H_\eps+g^2 N \varphi_\eps(0,0)$ converges to  a self-adjoint operator
$H_{\ren}$ as $\eps\downarrow 0$.

\subsubsection{Convergence of the renormalized action}
\label{222}

\bl{convexp2}
If $\alpha\in\RR$, then for every $x\in\BRN$,
\eq{warai2}
\lim_{\eps\downarrow 0}\Ebb^x[|\e^{\alpha U_\eps^T}-\e^{\alpha U_0^T}|]=0,\quad
U=S^{\od}, X, Y, Z.
\en
\el
\proof
Let $U=X$. We obtain that $|X_\eps^T|\leq \int_{-T}^{T} V_{\rm C}(B_s)ds$, where $\d V_{\rm C}(x)=
C\ij {|x^i-x^j|}^{-1}$ is Coulomb potential with a constant $C$, and the fact that
$$
\Ebb^x\!\left[|\e^{\alpha X_\eps^T}-\e^{\alpha X_0^T}|\right]
\leq
2 \Ebb^x\!\left[\e^{\alpha \int_{-T}^T V_{\rm C}(B_s)ds}\right]<\infty.
$$
Since $X_\eps^T\to X_0^T$ a.s. with respect to $\IP^x$ for every $x\in\BRN$,  the Lebesgue dominated 
convergence theorem implies
\kak{warai2}.

Let $U=Y$. It suffices to show that $\sup_{x\in\BRN}\Ebb^x\!\left[|\e^{\alpha (Y_\eps^T-Y_0^T)}-1|\right]
\to 0$. We have
\eq{ilo}
\Ebb^x\!\left[\lk \e^{\alpha (Y_\eps^T-Y_0^T)}-1\rk^2 \right]=
\Ebb^x\!\left[\e^{2\alpha (Y_\eps^T-Y_0^T)}\right]+
1-2\Ebb^x\!\left[\e^{\alpha (Y_\eps^T-Y_0^T)}\right].
\en
We will show below that $\lim_{\eps\downarrow 0} \Ebb^x[\e^{\alpha (Y_\eps^T-Y_0^T)}]=1$.
Define the random process $\delta \Phi_t=\Phi_{\eps,t}-\Phi_{0,t}$ so that
$$
Y_\eps^T-Y_0^T=\int_{-T}^{T} \delta\Phi_t \cdot dB_t.
$$
By the Girsanov theorem
\eq{gir}
1=\Ebb^x[\e^{\alpha \int_{0}^ {2T} \delta\Phi_t \cdot dB_t -\frac{\alpha^2}{2} \int_{0}^ {2T} |\delta\Phi_t|^2 dt}]
\en
for every $\alpha\in\RR$, hence it follows that
\eq{result}
\lk \Ebb^x\!\left[\e^{\alpha (Y_\eps^T-Y_0^T)} \right]-1\rk^2 \leq
\Ebb^x\!\left[\e^{2\alpha\int_{-T}^{T} \delta\Phi_t \cdot dB_t}\right]
\Ebb^x\!\left[\lk1-\e^{-\frac{\alpha^2}{2}\int_{-T}^{T} |\delta\Phi_t|^2 dt}\rk^2\right].
\en
We see that by \kak{gir} again
\eq{bound}
\sup_{x\in\BRN}\Ebb^x\!\left[\e^{2\alpha\int_{-T}^{T} \delta\Phi_t \cdot dB_t}\right]
\leq
\sup_{x\in\BRN}\lk \Ebb^x\!\left[\e^{4\alpha^2\int_{-T}^{T} |\delta\Phi_t|^2 dt}\right]\rk ^\han
\en
and furthermore
\eq{233}
\Ebb^x\!\left[\lk1-\e^{-\frac{\alpha^2}{2}\int_{-T}^{T} |\delta\Phi_t|^2 dt}\rk^2\right]
\leq
\Ebb^x\!\left[\left|\frac{\alpha^2}{2}\int_{-T}^{T} |\delta\Phi_t|^2dt\right|^2\right]\to 0
\en
as $\eps\downarrow 0$. Here \kak{233}  can be shown by Lemma \ref{prop:convergence}. The right-hand
side of \kak{bound} is uniformly bounded in $\eps$, which can be proven  in the same way as in the
proof of Lemma \ref{prop:exp-int}. Hence \kak{result} converges to zero as $\eps\downarrow 0$ and
\kak{warai2} for $U=Y$ follows.

Let $U=Z$. It suffices to show that $\sup_{x\in\BRN}\Ebb^x\!\left[|\e^{\alpha (Z_\eps-Z_0)}-1|\right]\to 0$.
We have
\begin{align*}
Z_\eps-Z_0
&=
2\iijj \int_{-T}^T
\!\!\!
ds\int_\BR\!\!\!
{\e^{-ik\cdot(B_{[s+\tau]_T-s}^i+x^i-B_{[s+\tau]_T-s}^j-x^j)}\e^{-([s+\tau]_T-s)
\omega(k)}}\\
&\hspace{2cm}\times \frac{1}{\omega(k)}
\beta(k)\kp (1-\e^{-\eps|k|^2})dk.
\end{align*}
Let $\eta_\eps=\alpha (Z_\eps-Z_0)$.
It can be directly seen that
$
\d |\eta_\eps|^n\leq c^n \alpha^n T^n \eps^n
$
for a constant $c$. Then we have $\d \Ebb^x[\e^{\eta_\eps}]=
1+\sum_{n\geq 1}\frac{1}{n!} \Ebb^0[\eta_\eps^n]$ and
$$
\d \sum_{n\geq 1}\frac{1}{n!}  \Ebb^x[|\eta_\eps|^n] \leq \sum_{n\geq 1}\frac{1}{n!}c^n T^n \eps^n\to 0
$$
as $\eps\downarrow 0$, uniformly in $x\in\BRN$. Thus \kak{warai2} for $U=Z$ follows. For $U=S^{\od}$ we obtain
\kak{warai2} in a similar way.
\qed

\begin{lemma}
\label{54321}
Let $\alpha\in\RR$ and  $f,h\in L^2(\BRN)$. Then
\begin{eqnarray}
\lefteqn{
\lim_{\eps\downarrow 0}\int_{\BRN} dx  \Ebb^x [f(B_{-T}) h(B_T)  \e^{-\int_{-T}^TV(B_s)ds}
\e^{\alpha S_\eps^\ren }]\non} \\
&&\label{hanatarasi}
= \int_{\BRN} dx  \Ebb^x [f(B_{-T} )h(B_T)\e^{-\int_{-T}^TV(B_s)ds}\e^{\alpha S_0^\ren }].
\end{eqnarray}
\end{lemma}
\proof
Write ${S_\eps}=
S^{\od,T}_\eps+X_\eps^T+Y_\eps^T+Z_\eps^T$. Then by telescoping we have
that
\begin{align*}
&
\lefteqn{
\left|\int_{\BRN} dx   \Ebb^x \left[f(B_{-T}) h(B_T) \e^{-\int_{-T}^TV(B_s)ds}
\lk \e^{\alpha S_\eps^\ren } - \e^{\alpha S_0^\ren }\rk \right]\right| } \\
&\leq
\e^{2T\|V\|_\infty}\int_{\BRN} dx  |f(x)|\Ebb^x \left[ |h(B_{T})|
\lk \e^{\alpha S_\eps} - \e^{\alpha S_0}
\rk \right]\\
&\leq
\e^{2T\|V\|_\infty}\int_{\BRN} dx  |f(x)|
\lk
\Ebb^x
\left[ |h(B_{T})|^2\right]\rk^{\han}
E_\eps(x),
\end{align*}
where $E_\eps(x)= \lk \Ebb^x \left[\lk \e^{\alpha S_\eps} - \e^{\alpha S_0}\rk^2\right]\rk^\han$.
Note that by reasoning like in Lemma \ref{prop:exp-int} we can show that $\sup_{x\in\BRN}E_\eps(x)
<\infty$, and by Lemma \ref{convexp2}  we have that $\lim_{\eps\downarrow 0}E_\eps(x)=0$ for every 
$x\in\BR$. Hence the Lebesgue dominated convergence theorem implies that the second term converges 
to zero, and the lemma follows.
\qed

\bl{maindenai}
It follows that
\eq{sasa21}
\lim_{\eps\downarrow 0} (f\otimes \one,\e^{-2T(\hn_\eps+g^2N\varphi_\eps(0,0))} h\otimes  \one)
= \int_{\BRN} \Ebb^x\!\left[\ov{f(B_{-T})} h(B_T)\e^{-\int_{-T}^T V(B_s) ds} \e^{\frac{g^2}{2} S_0^\ren }
\right]dx,
\en
where
\begin{align}\label{hiroo}
S_0^\ren
&=
2\ij\int_{-T}^T \varphi_0(B_s^i-B_s^j,0)ds
+2\iijj \int_{-T}^T ds\int_s^T\nabla\varphi_0(B_t^i-B_s^j,t-s) \cdot dB_t  \non \\
&
\qquad -2\iijj \int_{-T}^T \varphi_0(B_T^i-B_s^j,T-s)ds,
\end{align}
with the integrands given by $\varphi_0(X,t)=\int_\BR\frac{e^{-ikX}e^{-|t|\omega(k)}}{2\omega(k)}
\beta(k) \kp dk$ and $\nabla \varphi_0(X,t)=\int_\BR\frac{-ik e^{-ikX}e^{-|t|\omega(k)}}{2\omega(k)}
\beta(k) \kp dk$.
\el
\proof
We have
\eq{554321}
(f\otimes \one, \e^{-2T(\hn_\eps+g^2N\varphi_\eps(0,0))}h\otimes  \one) =
\int_{\BR} \Ebb^x\!\left[\ov{f(B_{-T})} h(B_T)\e^{-\int_{-T}^T V(B_s) ds}
\e^{\frac{g^2}{2} S_\eps^\ren } \right]dx .
\en
The right-hand  side above converges to $\int_{\BR} \Ebb^x [\ov{f(B_{-T})} h(B_T)
\e^{-\int_{-T}^T V(B_s) ds}\e^{\frac{g^2}{2} S_0^\ren} ]dx$ as $\eps\downarrow 0$.
Thus \kak{sasa21} follows. We also see that
\begin{align}
S_0^\ren
&=
2 \ij \int_{-T}^T \varphi_0(B_s^i-B_s^j,0) ds
+ 2\iijj \int_{-T}^T ds\int_s^{[s+\tau]_T}\nabla\varphi_0 (B_t^i-B_s^j,t-s) \cdot dB_t\non\\
&
\qquad - 2\iijj \int_{-T}^T \varphi_0(B^i_{[s+\tau]_T}-B^j_s,[s+\tau]_T-s) ds.
\end{align}
Taking $\tau=T$, we obtain \kak{hiroo}.
\qed

\subsubsection{Extension beyond the vacuum vector}
Now we extend the result in Section \ref{222} from vectors of the form $f\otimes \one$ to more general
vectors of the form $f\otimes F(\phi(f_1),\ldots,\phi(f_n))\one $, with $F\in \ms  S(\RR^n)$, where
$\phi(f)$ stands for a scalar field given by
$\frac{1}{\sqrt 2} (\add(\hat f)+a(\widetilde{\hat f}))$,  where $\widetilde{\hat f}(k)=\hat f(-k)$.
To do this we
need a Feynman-Kac-type formula giving a representation of $\e^{-2T H_\eps}$.

Denote
\eq{yoshida}
H_{-k}(\RR^n)=\{f\in \ms S_\RR'(\RR^n)\,|\,\widehat f\in L_{\rm loc}^1(\RR^n), |\cdot|^{-k/2}\widehat f\in
L^2(\RR^n)\}
\en
endowed with the norm $\d \|f\|_{H_{-k}(\RR^n)}^2=\int_{\RR^n} |\widehat f(x)|^2|x|^{-k} dx$. Recall that a
Euclidean field is a family of Gaussian random variables $\{\phi_{\rm E}(F), \, F\in H_{-1}(\RR^4)\}$ on a
probability space $(Q_{\rm E}, \Sigma_{\rm E}, \mu_{\rm E})$, such that the map $F \mapsto \phi_{\rm E}(F)$
is linear, and their mean and covariance are given by
$$
\Ebb_{\mu_{\rm E}}[\phi_{\rm E}(F)]=0 \quad \mbox{and} \quad \Ebb_{\mu_{\rm E}}[\phi_{\rm E}(F)
\phi_{\rm E}(G)]=\half (F,G)_{H_{-1}(\RR^4)}.
$$
For the reader's convenience we summarize in  Appendix \ref{appb} some basic facts on Euclidean fields and
their representation in $L^2$ space (including the operators $\JJ_t:\ffff \to L^2(Q_{\rm E};\mu_{\rm E})$),
which will be used here. In what follows, we identify $\hhh$ with the set of $\ffff $-valued $L^2$ functions
$L^2(\RR^{3N};\ffff )$, i.e., $F\in \hhh$ can  be regarded as a function  $\RR^{3N}\ni x\mapsto F(x)\in \ffff $
such that $\int_{\RR^{3N}}\|F(x)\|^2_\ffff  dx<\infty$.
\bp{FKF}
Let $F, G\in\hhh$. Then
\begin{align}
&(F, \e^{-2TH_\eps}G)\non\\
&\label{fkf}
=\int_{\BRN} dx  \Ebb^x \left[\e^{-\int_{-T}^T V(B_s) ds} \Ebb_{\mu_{\rm E}} \left[\JJ_{-T} F(B_{-T})\cdot
\e^{-\phi_{\rm E}(\int_{-T}^T\sum_{j=1}^N \delta_s\otimes \tilde\varphi(\cdot-B_s^j) ds)} \JJ_TG(B_T)\right]\right],
\end{align}
where $\tilde \varphi_\eps(x)=\lk \e^{-\eps|\cdot|^2/2}\one_\La^\perp /\sqrt{\omega}\rk^\vee{}(x)$, and $\delta_s(x) =
\delta(x-s)$ is Dirac delta distribution with mass on $s$.
\ep
\proof
See \cite[Theorem 6.3]{lhb11}.
\qed

\bl{h123}
Let $ \rho_j\in H_{-\han}(\BR)$ for $j=1,2$, $f,h\in\LRN$ and $\alpha,\beta\in\CC$. Then
\begin{eqnarray}
\lefteqn{
\lim_{\eps\downarrow0} (f\otimes \e^{\alpha\phi(\rho_1)}\one,
\e^{-2T(\hn_\eps+g^2N\varphi_\eps(0,0))} h\otimes  \e^{\beta\phi(\rho_2)}\one) } \non\\
&&\label{main22}
= \int_{\BRN} \Ebb^x\!\left[\ov{f(B_{-T})} h(B_T)\e^{-\int_{-T}^T V(B_s) ds} \e^{\frac{g^2}{2}
S_0^\ren +\frac{1}{4}\xi} \right]dx,
\end{eqnarray}
where
\begin{align*}
\xi=\xi(g)
&=
\bar \alpha^2\|\rho_1/\sqrt\omega \|^2+\beta^2\|\rho_2/\sqrt\omega\|^2+2\bar
\alpha \beta (\rho_1/\sqrt\omega, \e^{-2T\omega} \rho_2/\sqrt\omega) \\
&
\qquad + 2\bar \alpha g \sum_{j=1}^N
\int_{-T}^Tds \int_\BR dk \frac{\widehat \rho_1(k)}{\sqrt{\omega(k)}}\kp
\e^{-|s-T|\omega(k)}\e^{-ikB_s^j} \\
&
\qquad
+ 2\beta  g
\sum_{j=1}^N
\int_{-T}^Tds \int_\BR dk \frac{\widehat \rho_2(k)}{\sqrt{\omega(k)}}\kp
\e^{-|s+T|\omega(k)}\e^{-ikB_s^j}.
\end{align*}
\el
\proof
By the functional integral representation \kak{fkf} we have
\begin{align*}
&
(f\otimes \e^{\alpha\phi(\rho_1)}\one,
\e^{-2T(\hn_\eps+g^2N \varphi_\eps(0,0))} h\otimes
\e^{\beta\phi(\rho_2)}\one)
=
\int_{\BRN} \!\!\!\!\!  dx  \Ebb^x\!\left[\bar f(B_{-T} )h(B_T) \e^{-\int_{-T}^TV(B_s) ds}\right.\\
&\times
\left.
\Ebb_{\mu_E}[\e^{\bar \alpha\phi_E(\delta_{-T}\otimes \rho_1)}
\e^{\beta \phi_E(\delta_{T}\otimes \rho_2)} \e^{g \phi_E(-\sum_{j=1}^N
\int_{-T}^ T \delta_s\otimes  \tilde \varphi_\eps (\cdot-B_s^j)ds)} ]\right]
\e^{-2Tg^2 N\varphi_\eps(0,0)}.
\end{align*}
It can be
directly seen that
$$
\Ebb_{\mu_E}\!\!\left[\e^{\bar \alpha\phi_E(\delta_{-T}\otimes \rho_1)}
\e^{\beta \phi_E(\delta_{T}\otimes \rho_2)}\e^{g \phi_E(-\sum_{j=1}^N
\int_{-T}^ T \delta_s\otimes  \tilde \varphi_\eps (\cdot-B_s^j)ds)} \right]
\e^{-2Tg^2 N\varphi_\eps(0,0)} =\e^{\frac{g^2}{2}S_\eps^\ren +\frac{1}{4}\xi_\eps},
$$
where $\xi_\eps$ is defined by $\xi$ with $\kp$ replaced by $\kp \e^{-\eps|k|^2/2}$. Thus
\begin{align*}
& (f\otimes \e^{\alpha\phi(\rho_1)}\one, \e^{-2T(\hn_\eps+g^2N\varphi_\eps(0,0))}
h\otimes  \e^{\beta\phi(\rho_2)}\one) \\
&
= \int_{\BRN} \!\!\! dx \Ebb^x\!\left[\ov{f(B_{-T})} h(B_T)\e^{-\int_{-T}^T V(B_s) ds} \e^{\frac{g^2}{2}
S_\eps^\ren +\frac{1}{4}\xi_\eps} \right].
\end{align*}
Notice that with a constant $C$ we have $\xi_\eps\leq C$ uniformly in the paths and $\eps\geq0$. Hence
we can complete the proof of the lemma in a similar way to Lemma \ref{maindenai}.
\qed
Consider the dense subspace  $\D$ of  $\hhh$ given by
\begin{align*}
&\D={\rm L.H.}\, \{f\otimes \one \,|\, f\in\LRN\} \, \cup \\
&
\lkk f\otimes F(\phi(f_1),\ldots,\phi(f_n)) \,|\, F\in \ms S(\RR^n),
f_j\in C_0^\infty(\BR), 1\leq j\leq n,
 f\in\LRN \rkk.
\end{align*}
By Lemma \ref{h123} the next result is immediate.
\bl{h2}
Let $\Phi=f\otimes F(\phi(u_1),\ldots,\phi(u_n))$, $\Psi=h\otimes G(\phi(v_1),\ldots,\phi(v_m)) \in \D$.
Then
\begin{align}
\lim_{\eps\downarrow0}
(\Phi, \e^{-2T (\hn_\eps+g^2 N\varphi_\eps(0,0))} \Psi)
&=
(2\pi)^{-(n+m)/2} \int_{\RR^{n+m}}dK_1 dK_2 \ov{\widehat F(K_1)} \widehat G(K_2) \non \\
& \hspace{-1cm}\times
\int_{\BRN} dx   \Ebb^x\!\left[\ov{f(B_{-T})} h(B_T) \e^{-\int_{-T}^T V(B_s) ds} \e^{\frac{g^2}{2}
S_0^\ren + \frac{1}{4} \xi(K_1,K_2)} \right],
\label{310}
\end{align}
where
\begin{align*}
\xi(K_1,K_2)
&=
-\|K_1\cdot u/\sqrt\omega\|^2 -\|K_2\cdot v /\sqrt\omega\|^2 -
2 (K_1 \cdot u/\sqrt\omega, \e^{-2T\omega} K_2 \cdot v/\sqrt\omega)\\
& \qquad
- 2i g \sum_{j=1}^N \int_{-T}^Tds \int_\BR dk
\frac{K_1\cdot \widehat u(k)}{\sqrt{\omega(k)}}\kp \e^{-|s-T|\omega(k)}\e^{-ikB_s^j} \\
&
\qquad + 2i g \sum_{j=1}^N \int_{-T}^Tds \int_\BR dk
\frac{K_2\cdot \widehat v(k)}{\sqrt{\omega(k)}}\kp \e^{-|s+T|\omega(k)}\e^{-ikB_s^j}
\end{align*}
and $u=(u_1,...,u_n)$, $v=(v_1,...,v_m)$.
\el
\proof
Notice that $F(\phi(f_1),\ldots,\phi(f_n))
=(2\pi)^{-n/2} \int_{\RR^n} \widehat F(K)
\e^{i\phi (K\cdot f)} dK$.
Hence
\begin{align*}
& (\Phi, \e^{-2T(\hn_\eps+g^2 N \varphi_\eps(0,0))} \Psi)
= \frac{1}{(2\pi)^{(n+m)/2}}
\int_{\RR^{m+n}}
dK_1dK_2\ov{\widehat F(K_1)} \widehat G(K_2)\\
&\hspace{3cm}\times (f\otimes \e^{-i\phi(K_1\cdot f)},
\e^{-2T(\hn_\eps+g^2 N \varphi_\eps(0,0))}
h\otimes \e^{-i\phi(K_2\cdot h)}).
\end{align*}
Thus the statement follows from Lemma \ref{h123}.
\qed

\subsubsection{Uniform lower bound}
\label{223}
In this section we show a crucial lower bound on $H_\eps+g^2 N \varphi_\eps(0,0)$ uniform in
$\eps>0$, and give the proof of Theorem \ref{main}.

\bc{nelson}
There exists $C \in \RR$ such that $\hn_\eps+g^2 N\varphi_\eps(0,0)>C$, uniformly in $\eps>0$.
\ec
\proof
 Consider the function
$$
W(x^1,...,x^N)=\sum_{j=1}^N |x^j|^2.
$$
We denote $H_\eps$ with $V$ replaced by $\delta W$ by $H_\eps(\delta)$, for $\delta\geq0$.
Then $-\half \sum_{j=1}^N \Delta_j +\delta W$, $\delta>0$, has a compact resolvent, which implies
that $H_\eps(\delta)$ for $\delta>0$  has the unique ground state $\gr(\delta)$ by \cite{spo98,ger00},
see Remark \ref{existence} below. By the Feynman-Kac formula \kak{fkf} we see that $\e^{-TH_\eps(\delta)}$
is positivity improving for $\eps>0$, i.e., $(F, \e^{-TH_\eps(\delta)}G)>0$ for $F,G\in \hhh$ such that
$F, G\geq 0$. Hence it follows that $\gr(\delta)>0$. In particular, $(f\otimes \one, \gr(\delta))\neq 0$,
for every $0\leq  f\in\LRN$, where $f\not\equiv 0$. Thus
\eq{inf}
\is \left(H_\eps(\delta)+g^2 N\varphi_\eps(0,0)\right)
=
-\lim_{T\to \infty}\frac{1}{T}\log (f\otimes \one, \e^{-T(H_\eps(\delta)+g^2 N\varphi_\eps(0,0))}f\otimes\one),
\en
for every $0\leq f\in \LRN$. By Lemma \ref{prop:exp-int} there exists a constant $b$ such that
\begin{align*}
&(f\otimes \one, \e^{-2T(H_\eps(\delta)+g^2 N\varphi_\eps(0,0))}f\otimes \one)
=
\int_{\BRN} dx \Ebb^x[f(B_{-T}) f(B_{T})\e^{-\int_{-T}^{T}\delta W(B_s) ds}\e^{S_\eps^\ren}]\\
&\leq
\int_{\BRN} dx  \Ebb^x[|f(B_{-T})| |f(B_{T})| \e^{S_\eps^\ren}]\\
&\leq
\|f\|^2 \e^{b (1+2T)},
\end{align*}
which implies, together with \kak{inf},  that
\eq{delta3}
\is \left(H_\eps(\delta)+g^2 N\varphi_\eps(0,0)\right)+{b} \geq0,\quad \delta>0.
\en
Note that $b$ is independent of $\delta$. Thus
\eq{delta1}
|(F, \e^{-2T(H_\eps(\delta)+g^2 N\varphi_\eps(0,0))}G)|\leq \|F\| \|G\| \e^{2 b T}
\en
follows for every $\delta>0$. Let $F,G\in\hhh$. By the Feynman-Kac formula \kak{fkf} we have
\begin{align*}
&
(F, \e^{-2TH_\eps(\delta)}G)\\
&
=\idx \Ebb^x\!\left[\e^{-\int_{-T}^T \delta W(B_s) ds} \Ebb_{\mu_{\rm E}}\!
\left[\JJ_{-T} F(B_{-T})\cdot  \e^{-\phi_{\rm E}(\int_{-T}^T\sum_{j=1}^N \delta_s\otimes
\tilde\varphi(\cdot-B_s^j) ds)} \JJ_TG(B_T)\right]\right].
\end{align*}
The Lebesgue dominated convergence theorem furthermore implies
$$
\lim_{\delta\downarrow 0}(F, \e^{-2T(H_\eps(\delta)+g^2 N\varphi_\eps(0,0))}G)
= (F, \e^{-2T(H_\eps(0)+g^2 N\varphi_\eps(0,0))}G).
$$
Taking the limit $\delta\downarrow 0$ on both sides of \kak{delta1}, we have
\eq{delta2}
|(F, \e^{-2T(H_\eps(0)+g^2 N\varphi_\eps(0,0))}G)|\leq \|F\| \|G\| \e^{2 b T}.
\en
This implies that  \kak{delta3} also holds for $\delta=0$. Since $H_\eps=H_\eps(0)+V$ and $V$
is bounded, we obtain
$$
\is(H_\eps+g^2 N\varphi_\eps(0,0))+{b_5}+\|V\|_\infty \geq0.
$$
Setting $C=-{b}-\|V\|_\infty$ yields the corollary.
\qed
\br{existence}
\rm{
Let $\Sigma$ be the infimum of the essential spectrum of the self-ajoint operator $h=-\half \sum_{j=1}^N
\Delta_j +V$ in $L^2(\RR^{3N})$ and $E=\is(-\half\sum_{j=1}^N  \Delta_j +V)$. Denote by $\grp$ the ground 
state of $h$. Then it is known that $H_\eps$ has a ground state if and only if
$$
\d \lim_{T\to\infty}\frac{(\grp\otimes\one, e^{-TH_\eps}\grp\otimes\one)^2}{(\grp\otimes\one, 
e^{-2TH_\eps}\grp\otimes\one)}>0.
$$
This was shown in \cite{spo98} (see also \cite[Section~6]{lhb11}) by using functional integration. Then
$H_\eps$ has a unique ground state if
$$
\Sigma-E>\frac{N^2}{4} \int_\BR \e^{-\eps|k|^2}\beta(k) \kp dk
$$
(see also \cite[Theorem 6.6]{lhb11}). In particular, in the case of $V(x^1,...,x^N)=\delta \sum_{j=1}^N
|x^j|^2$, the operator $H_\eps$ has a unique ground state for every $\eps>0$ and $\delta>0$, since
$\Sigma-E=\infty$.
}
\er

Now we can complete the proof of the main theorem.

\medskip
\noindent
\emph{Proof of Theorem \ref{main}.}
Let $F, G\in \hhh$ and $C_\eps(F,G)=(F, \e^{-t(H_\eps+g^2N\varphi_\eps(0,0))}G)$. By Lemma \ref{h123} we
obtain that $C_\eps(F,G)$ is convergent as $\eps\downarrow0$, for every $F,G\in\D$. By the uniform bound
$$
\|\e^{-t(H_\eps+g^2N\varphi_\eps(0,0))}\|<\e^{-tC}
$$
obtained from Corollary  \ref{nelson} and since $\D$ is dense in $\hhh$, it follows that
$\{C_\eps(F,G)\}_\eps$ is a Cauchy sequence for $F,G\in\hhh$. Let $C_0(F,G)=\lim_{\eps\downarrow 0}C_\eps(F,G)$.
Hence we get $|C_0(F,G)|\leq \e^{-tC}\|F\|\|G\|$. The Riesz theorem implies that there exists a bounded operator
$T_t$ such that
$$
C_0(F,G)=(F, T_t G),\quad F,G\in \hhh.
$$
Thus $\slim_{\eps\downarrow 0}\e^{-t(\hn_\eps+g^2N\varphi_\eps(0,0))}=T_t$ follows. Furthermore, we also have
that
$$
\slim_{\eps\downarrow 0} \e^{-t(\hn_\eps+g^2N\varphi_\eps(0,0))}
\e^{-s(\hn_\eps+g^2N\varphi_\eps(0,0))}=\slim_{\eps\downarrow 0}
\e^{-(t+s)(\hn_\eps+g^2N\varphi_\eps(0,0))}=T_{t+s}.
$$
Since the left-hand side above is $T_t T_s$, the semigroup property of $T_t$ follows. Since
$\e^{-t(\hn_\eps+g^2N\varphi_\eps(0,0))}$ is a symmetric semigroup, $T_t$ is also symmetric. By
the functional integral representation \kak{310} the functional $(F, T_t G)$ is continuous at $t=0$
for every $F, G \in \D$. Given that $\D$ is in $\hhh$ and $\|T_t\|$ is uniformly bounded in a
neighborhood of  $t=0$, it also follows that  $T_t$ is strongly continuous at $t=0$. Then the
semigroup version of Stone's theorem \cite[Proposition 3.26]{lhb11} implies that there exists a
self-adjoint operator $\hni$, bounded from below, such that
$T_t=\e^{-t\hni}$, $t\geq0$.  The proof is
completed by setting $E_\eps=-g^2N\varphi_\eps(0,0)$.
\qed
We established the existence of the renormalized Hamiltonian $H_\ren$. We can obtain explicitly
the pair potential associated with $H_\ren$.
\bc{pair potential}
The pair potential associated with $\hni$ is given by $\frac{g^2}{2}S_0^\ren $.
\ec
\proof
By Lemma \ref{maindenai} we see that
\eq{sasa212}
(f\otimes \one, \e^{-2T\hni} h\otimes \one)
=\int_{\BRN} dx  \Ebb^x \left[\ov{f(B_{-T})} h(B_T)\e^{-\int_{-T}^T V(B_s) ds} \e^{\frac{g^2}{2} S_0^\ren}
\right].
\en
\qed

\section{Effective potential in the weak coupling limit}
In this section we consider the weak coupling limit of the renormalized Hamiltonian.
 In order to have a physically reasonable
effective potential, we take the dispersion relation
$$
\omega_\nu(k)=\sqrt{|k|^2+\nu^2}
$$
with positive mass $\nu>0$ instead of $\omega(k)=|k|$,  set the IR cutoff to zero and
take the cutoff function to be
$$
\vpe(k)= (2\pi)^{-3/2}\e^{-\eps|k|^2/2}.
$$
The Hamiltonian
is defined on $L^2(\RR^{3N})\otimes\ffff $ and given by
$$
\hn_\eps=\hp\otimes\one+\one\otimes\hf+\hi,
$$
where $\hp=\sum_{j=1}^N (-\half \Delta_j) +V(x_1,...,x_N)$ denotes the $N$-body Schr\"odinger operator, and
$$
\hf=\int_\BR \omega_\nu(k)\add(k) a(k)dk
$$
is the free massive boson field. We scale the Hamiltonian by replacing the annihilation operator $a$ and the
creation operator $\add$ by $\kappa a$ and $\kappa \add$, respectively, where $\kappa>0$ is the scaling
parameter. Then $\hn_\eps$ changes to
\eq{weak}
\hn_\eps(\kappa)=\hp\otimes \one +\kappa^2 \one\otimes\hf+\kappa \hi.
\en
This scaling also implies the transformations $\omega\mapsto \kappa^2 \omega$ and $\vp\mapsto \kappa^2\vp$,
while the energy renormalization term scales as
\eq{energy}
E_\eps(\kappa)=-g^2 N  \int_\BR \frac{\e^{-\eps |k|^2}}{2(2\pi)^3 \omega_\nu(k)}
\frac{\kappa^2}{\kappa^2\omega_\nu(k)+|k|^2/2} dk.
\en
By Theorem \ref{main} there exists
a self-adjoint operator $H_{\ren}(\kappa)$ such that
\eq{main2}
\lim_{\eps\downarrow 0} (f\otimes\one, \e^{-t(H_\eps(\kappa)-E_\eps(\kappa))}
h\otimes  \one)=(f\otimes\one, \e^{-tH_{\ren}(\kappa)}h\otimes \one).
\en
The next proposition is established in \cite{dav79,hir99}.
\bp{wcl}
We have
$$
\slim_{\eps\downarrow 0} \lim_{\kappa\to\infty} \e^{-t(\hn_\eps(\kappa)-E_\eps(\kappa))}=
\e^{-th_{\rm eff}}\otimes P_\Omega,
$$
where $P_\Omega$ denotes the projection to $\{z\one\,|\,z\in\CC\}\subset\ffff $ and
$$
h_{\rm eff}=-\half \sum_{j=1}^N \Delta_j + V(x^1,...,x^N)-\frac{g^2}{4\pi }\sum_{i<j}
\frac{\e^{-\nu|x_i-x_j|}}{|x_i-x_j|}.
$$
\ep

We are now interested in the scaling limit of $H_{\ren}(\kappa)$ when $\kappa\to\infty$. By Theorem
\ref{main} we see that
\bl{main4}
If $f,h\in L^2(\BRN)$, then
\eq{weakcouplinglimit}
\lim_{\kappa\to\infty}(f\otimes \one, \e^{-t H_\ren (\kappa)}h\otimes \one) =(f,\e^{-t h_{\rm eff}} h).
\en
\el
\proof
By Lemma \ref{h123} we have
\eq{sasa2121}
(f\otimes \one, \e^{-2T H_\ren (\kappa)}h\otimes \one) =\int_{\BRN} dx
\Ebb^x \! \left[\ov{f(B_{-T})} h(B_T)\e^{-\int_{-T}^T V(B_s) ds} \e^{\frac{g^2}{2} S_0^\ren (\kappa)}\right],
\en
where
\begin{align}\label{hiro}
S_0^\ren (\kappa)
&
= 2\ij\int_{-T}^T \varphi_0(B_s^i-B_s^j,0,\kappa)ds
+2\iijj \int_{-T}^T \!\!\!\!\!
ds\int_s^T
\!\!\!\!\!  \nabla\varphi_0(B_t-B_s,t-s,\kappa) \cdot dB_t\non\\
&
\qquad -2\iijj \int_{-T}^T \varphi_0(B_T-B_s,T-s,\kappa)ds,
\end{align}
and
\eq{sasa33}
\varphi_0(x,t,\kappa)=\frac{1}{(2\pi)^3}\int_{\BR} \frac{\e^{-ik\cdot x} \e^{-\kappa^2\omega(k)|t|}}{2\omega(k)}
\frac{\kappa^2}{\kappa^2 \omega(k)+|k|^2/2}  \kp dk.
\en
In particular, for $t=0$ we have
$$
g^2\ij \varphi_0(x^i-x^j,0,\kappa)ds \;\to\; \frac{g^2}{4\pi} \sum_{i<j}\frac{\e^{-\nu|x^i-x^j|}} {|x^i-x^j|},
$$
and for $t\not=0$,
$$
|\nabla\varphi_0(X, t, \kappa)|\to 0, \quad |\varphi_0(X, t, \kappa)|\to 0
$$
pointwise as $\kappa\to\infty$. It can be shown in the same way as in Lemma \ref{maindenai} that
\begin{align*}
&
\lim_{\kappa\to\infty} \int_{\BRN} dx
\Ebb^x \! \left[\ov{f(B_{-T})} h(B_T)\e^{-\int_{-T}^T V(B_s) ds} \e^{\frac{g^2}{2} S_0^\ren (\kappa)}\right]\\
&=
\int_{\BRN} dx \Ebb^x\!\left[\ov{f(B_{-T})} h(B_T)\e^{-\int_{-T}^T V(B_s) ds} \e^{\frac{g^2}{4\pi}
\sum_{i<j} \int_{-T}^T \frac{\e^{-\nu|B^i_s-B^j_s|}} {|B^i_s-B^j_s|}ds} \right].
\end{align*}
This completes the proof of the corollary.
\qed
\bc{main5}
If $F, G\in\ms D$, then
\eq{weakcouplinglimit2}
\lim_{\kappa\to\infty}(F,  \e^{-t H_\ren (\kappa)} G) = (F,(\e^{-t h_{\rm eff}}\otimes P_\Omega)  G).
\en
\ec
\proof
This follows from Lemmas \ref{h123} and  \ref{main4}.
\qed

\appendix
\section{Kato-class potentials}
\label{appa}
We say that a potential $V: \RR^d \to \RR$ belongs to the Kato-class relative to the Laplacian, whenever
\eq{kato2}
\lim_{t\downarrow 0}\sup_{x\in\RR^d} \Ebb \!\left [\int_0^t |V(W^x_s)|ds\right ]=0
\en
where $(W^x_t)_{t\ge 0}$ is a standard $d$-dimensional Brownian motion starting at $x\in\RR^d$.
We will denote by $\ms K_d$ the set of all such potentials.  For details
on Kato-class potentials we refer to \cite[Chapter 3.3]{lhb11} and \cite{as82, cfks08}.

\bp{multi}
If $V\in \ms K_d$, then $W(x)=\sum_{i\not =j}^N V(x^i-x^j) \in \ms K_{dN}$, with the notation $x=(x^1,...,x^N)
\in \RR^{dN}$.
\ep
For a proof we refer e.g. to \cite[p.7]{cfks08}. An equivalent characterization of Kato-class potentials is
as follows. A potential $V\in \ms K_d$ if and only if
\eq{eq}
\lim_{r\downarrow 0}\sup_{x\in\RR^d}\int_{|x-y|<r}|g(x-y) V(y)|dy=0 \quad \mbox{with} \quad
g(x)=\lkk \begin{array}{ll}
|x|& d=1\\
-\log|x|&d=2\\
|x|^{2-d}&d\geq 3.
\end{array}
\right.
\en
Examples of Kato-class potentials include
(1)  $|x|^{-(2-\eps)}$ with  $d=3$ for any $\eps>0$,
(2) $V\in L^p(\RR^d)+L^\infty(\RR^d)$ with $p=1$ for $d=1$, and $p>d/2$ for $d\geq 2$.
It is also known that the function $\e^{\int_0^t V(W^x_s) ds}$ of the $d$-dimentional Brownian motion $(W^x_t)_{t\ge 0}$ is integrable if $V$ is Kato-class.
\bp{expexp}
Let $0\leq V\in \ms K_d$. Then there exist $\beta, \gamma>0$ such that
\eq{alpha}
\sup_{x\in\RR^d}\Ebb[\e^{\int_0^t V(W^x_s) ds}]\le \gamma \e^{t\beta}.
\en
\ep
\proof
See \cite[Lemma 3.38]{lhb11}.
\qed

\section{Schr\"odinger representation and Euclidean field}
\label{appb}
In this section  Hilbert spaces $H_{-\han}(\RR^3)$ and
$H_{-1}(\RR^4)$
are  given by \kak{yoshida}.
It is well known that the boson Fock space $\ffff $ is unitarily equivalent to $L^2(Q,\mu)$, where this
space consists of square integrable functions on a probability space $(Q,\Sigma,\mu)$. Consider the
family of Gaussian random variables $\{\phi_0(f), \, f\in H_{-1/2}(\BR)\}$ on $(Q,\Sigma,\mu)$ such
that $\phi_0(f)$ is linear in
$f\in H_{-\han}(\RR^3)$,  and their mean and covariance are given by
$$
\Ebb_{\mu}[\phi_0(f)]=0 \quad \mbox{and} \quad \Ebb_{\mu}[\phi_0(f)\phi_0(g)]
=\half (f, g)_{H_{-\han}(\RR^3)}.
$$
Given this space, the Fock vacuum $\one_\ffff $ is unitary equivalent to $\one_{L^2(Q)}\in L^2(Q)$, and
the scalar field $\phi(f)$ is unitary equivalent to $\phi_0(f)$ as operators, i.e., $\phi_0(f)$ is
regarded as multiplication by $\phi_0(f)$. Then the linear hull of the vectors given by the Wick products
$:\prod_{j=1}^n \phi_0(f_j):$ is dense in $L^2(Q)$, where recall that Wick product is recursively defined
by
\begin{eqnarray*}
&&
{:\phi_0(f):} \; =\phi_0(f)  \\
&&
{:\phi_0(f)\prod_{j=1}^n \phi_0(f_j):} \: =
\phi_0(f):\prod_{j=1}^n \phi_0(f_j):-\half\sum_{i=1}^n (f,f_i)_{H_{-\han}(\RR^3)} :\prod_{j\not=i}^n \phi_0(f_j):
\end{eqnarray*}
This allows to identify $\ffff $ and $L^2(Q)$, which we have done in \kak{fkf}, i.e., $F\in \hhh$ can be regarded
as a function $\RR^{3N}\ni x\mapsto F(x)\in L^2(Q)$ such that $\int_{\RR^{3N}}\|F(x)\|^2_{L^2(Q)} dx<\infty$.

To construct a Feynman-Kac-type representation we use a Euclidean field. Consider the family of Gaussian random
variables $\{\phi_{\rm E}(F), \, F\in H_{-1}(\RR^4)\}$ with mean and covariance
$$
\Ebb_{\mu_E}[\phi_E(F)]=0 \quad \mbox{and} \quad \Ebb_{\mu_E}[\phi_E(F)\phi_E(G)]=\half (F, G)_{H_{-1}(\RR^4)}
$$
on a chosen probability space $(Q_E, \Sigma_E, \mu_E)$. Note that for $f\in H_{-1/2}(\BR)$ the relations
$$
\delta_t\otimes f\in H_{-1}(\RR^4) \quad  \mbox{and} \quad \|\delta_t\otimes f\|_{H_{-1}(\RR^4)} =
\|f\|_{H_{-1/2}(\BR)}
$$
hold, where $\delta_t(x) = \delta(x-t)$ is Dirac delta distribution with mass on $t$. The family of identities
used in \kak{fkf}
is  then given by $\JJ_t: L^2(Q)\to L^2(Q_{\rm E})$, $t\in\RR$, defined by the relations
$$
\JJ_t \one_{L^2(Q)}=\one_{L^2(Q_{\rm E})} \quad \mbox{and} \quad
\JJ_t {:\prod_{j=1}^m \phi(f_j):} \;=\; {:\prod_{j=1}^m \phi_{\rm E}(\delta_t\otimes f_j):}
$$
Under the identification
$\ffff \cong L^2(Q)$ it follows that
$$(\JJ_t F, \JJ_s G)_{L^2(Q_{\rm E})}=
(F, \e^{-|t-s|\hf }G)_\ffff $$ for $F, G\in\ffff $. For an extensive discussion of the details we refer to
\cite[Chapter 5]{lhb11}.

\bigskip
\noindent
{\bf Acknowledgments:}
FH acknowledges support of Grant-in-Aid for Science Research (B) 20340032 from JSPS, and thanks the 
hospitality of Universit\'e d'Aix-Marseille-Luminy, Universit\'e de Rennes 1 and Mittag-Leffler 
Institute, where part of this work has been  done. JL thanks IHES, Bures-sur-Yvette, for a visiting 
fellowship, Institut Mittag-Leffler, Stockholm, for the opportunity to organise the research-in-peace 
workshop ``Lieb-Thirring-type bounds for a class of Feller processes perturbed by a potential" during 
the period 25 July~--~9 August 2013, and gratefully thanks his colleague Fumio Hiroshima for an 
invitation to the workshop \emph{Quantum Field Theory and Related Topics}, held at RIMS Kyoto in November 
2012.

{

}


\begin{thebibliography}{99}
\bibitem[AS82]{as82}
M.  Aizenman and B. Simon,
Brownian motion and Harnack's inequality for Schr\"odinger Hamiltonians,
{\it Commun. Pure. Appl. Math. }{\bf 35} (1982), 209--273.

\bibitem[Ara01]{ara01}
A.  Arai,
Ground state of the massless Nelson model without infrared cutoff in a non-Fock representation,
{\it Rev. Math. Phys.} {\bf 13}(2001), 1075--1094.

\bibitem[BFS98]{bfs98}
V.  Bach,   J.  Fr\"ohlich and    I.  M.  Sigal,
Quantum electrodynamics of confined non-relativistic particles, {\it  Adv. Math. } {\bf 137}
(1998),   299--395.

\bibitem[BH09]{bh09}
V. Betz and F. Hiroshima,
Gibbs measures with double stochastic integrals on path space,
{\it Infinite Dimensional Analysis,  Quantum Probability and Related Topics}
{\bf 12} (2009),  135--152.


 \bibitem[BHLMS02]{bhlms02}
V.  Betz, F.  Hiroshima,  J.  L\H orinczi,
R.  A.  Minlos and H.  Spohn,
Ground state properties of the Nelson Hamiltonian --- A Gibbs measure-based approach,
{\it Rev. Math. Phys.} {\bf 14}  (2002), 173--198.


\bibitem[CFKS08]{cfks08}
H. L.  Cycon, R.  G.  Frose, W.  Kirsch
and B.  Simon,
{\it Schr\"odinger Operators}, 2nd ed., Springer, 2008.

\bibitem[Dav79]{dav79}
E. B.  Davies,   Particle-boson interactions and weak coupling limit,
{\it  J. Math. Phys.} {\bf 20}  (1979),  345--351.


\bibitem[FG02]{FG1}
F. Flandoli and M. Gubinelli. 
\newblock The {G}ibbs ensemble of a vortex filament. 
\newblock{\em Probab. Theory Related Fields}, 122(3):317--340, 2002.

\bibitem[Ger00]{ger00}
C.  G\'erard,
On the existence of ground states for massless Pauli-Fierz Hamiltonians,
{\it Ann.  Henri  Poincar\'e} {\bf 1}  (2000),  443--459.


\bibitem[GHPS12]{ghps12}
C. G\'erard, F. Hiroshima, A. Panati and A. Suzuki,
Removal of UV cutoff for the Nelson model with variable coefficients,
{\it Lett. Math. Phys.} {\bf 101} (2012), 305--322.

\bibitem[GL09]{GL}
M. Gubinelli and J. L\H{o}rinczi, 
Gibbs measures on Brownian currents, \emph{Commun. Pure Appl. Math.} \textbf{62}, 1-56, 2009.


\bibitem[Hir06]{hir06}
M. Hirokawa,
Infrared catastrophe for Nelson's model --- non-existence of ground state and soft-boson
divergence, {\it Publ. RIMS, Kyoto Univ.}  \textbf{42},  (2006), 897--922.


\bibitem[HHS05]{hhs05}
M. Hirokawa, F. Hiroshima and H. Spohn,
Ground state for point particles interacting through a massless scalar bose field,
{\it Adv. Math.} {\bf 191} (2005), 339--392.

\bibitem[Hir99]{hir99}
F. Hiroshima,
Weak coupling limit removing an ultraviolet cut-off for a Hamiltonian of
particles interacting with a quantized scalar field, {\it  J. Math. Phys.} {\bf  40}
(1999), 1215--1236.


\bibitem[LeG85]{LeG1}  J.F. le Gall, Sur le temps local d'intersection du
mouvement Brownien plan, et la m\'{e}thode de renormalisation de Varadhan,
\textit{S\'{e}m. de Prob. XIX}, 1983/84, LNM 1123, Springer-Verlag, Berlin
1985, 314-331.


\bibitem[LeG94]{LeG3}  J.F. le Gall, Exponential moments for the
  renormalized self-intersection local time of planar Brownian motion,
\textit{S\'{e}m. de Prob. XXVIII}, 1994, LNM 1583, Springer-Verlag, Berlin
1994, 172-180.

\bibitem[LHB11]{lhb11}
J. L\H orinczi,  F. Hiroshima and V. Betz,
{\it Feynman-Kac-Type Theorems and Gibbs Measures on Path Space}, de Gruyter Studies
in Mathematics {\bf 34},  2011.

\bibitem[LMS02]{lms02}
J. L\H{o}rinczi, R.A. Minlos and H. Spohn,
The infrared behavior in Nelson's model of a quantum particle coupled to a massless scalar
field, {\it Ann. Henri Poincar\'e}  {\bf 3} (2002), 1--28.

\bibitem[Nel64a]{nel64}
E. Nelson,
Interaction of nonrelativistic particles with a quantized scalar field,
{\it J.   Math.   Phys. } {\bf  5}  (1964),   1190--1197.


\bibitem[Nel64b]{nel64b}
E. Nelson,
Schr\"odinger particles interacting with a quantized scalar field, in: {\it Proc. Conference
on Analysis in Function Space}, W. T.  Martin and I. Segal (eds.), p. 87, MIT Press, 1964.

\bibitem[Sas05]{sas05}
I. Sasaki,
Ground state of the massless Nelson model in a non-Fock representation,
{\it J. Math. Phys.} {\bf 46} (2005), 102107.

\bibitem[Spo87]{spo87}
H. Spohn,
Effective mass of the polaron: a functional integral approach, {\it Ann. Phys.} {\bf 175}
(1987), 278--318.

\bibitem[Spo98]{spo98}
H.  Spohn,
Ground state of quantum particle coupled to a scalar boson field,
{\it  Lett.   Math.   Phys.  } {\bf 44}   (1998),   9--16.


\bibitem[Yo85a]{Yo1}  M. Yor, Compl\'{e}ments aux formules de Tanaka-Rosen, 
\textit{S\'{e}m. de Prob. XIX}, 1983/84, LNM 1123, Springer-Verlag, Berlin 1985,
332-348.

\bibitem[Yo85b]{Yo2}  M. Yor, Renormalisation et convergence en loi pour le temps
locaux d'intersection du mouvement brownien dans $\mathbf{R}^{3}$, \textit{
S\'{e}m. de Prob. XIX}, 1983/84, LNM 1123, Springer-Verlag, Berlin 1985,
350-365.

\bibitem[Yo86]{Yo3}
M.~Yor.
\newblock {Sur la repr\'esentation comme int\'egrales stochastiques des temps
  d'occupation du mouvement {B}rownien dans {${\bf R}\sp d$}}.
\newblock \emph{S\'em. de Prob. XX}, 1984/85,  LNM 1204,  Springer, Berlin, 1986, 
543--552.


\end{thebibliography}
\end{document}